\documentclass[fleqn,twoside,twocolumn,nofootinbib,showkeys]{revtex4} % Specifies the document class %,unsortedaddress
\usepackage[sec,nocpr]{ujp} % \usepackage[cyr]{ujp} for cyrillic

\def\bd{\begin{displaymath}}\def\ed{\end{displaymath}}
\def\be{\begin{equation}}\def\ee{\end{equation}}
\def\bea{\begin{eqnarray}}\def\eea{\end{eqnarray}}
\def\ba{\begin{array}}\def\ea{\end{array}}
\def\bs{\begin{split}}\def\es{\end{split}}
\def\nn{\nonumber}\def\lb{\label}

\def\a{\alpha}\def\b{\beta}\def\c{\chi}\def\d{\delta}
\def\f{\phi}
\def\l{\lambda}\def\m{\mu}\def\n{\nu}\def\q{\psi}\def\r{\rho}\def\s{\sigma}
\def\y{\eta}\def\x{\xi}\def\z{\zeta}

\def\D{\Delta}

\def\de{\partial}
\def\id{\equiv}\def\mo{{-1}}\def\ha{{1\over 2}}

\def\bdot{\!\cdot\!}

\def\diag{{\rm diag}}

\def\mn{{\mu\nu}}

\def\coo{coordinates }

\def\cc{coupling constant }

\def\sch{Schwarzschild }\def\ads{anti-de Sitter }
\def\poi{Poincar\'e }
\def\des{de Sitter }

\def\cor{commutation relations }

\def\cP{{\cal P}}\def\cQ{{\cal Q}}\def\cK{{\cal K}}\def\cJ{{\cal J}}\def\cD{{\cal D}}\def\cG{{\cal G}}\def\cH{{\cal H}}
\def\rad{\sqrt{1+\b p^2}}

\begin{document}
\title[The Snyder model and Quantum Field Theory]
{THE SNYDER MODEL AND QUANTUM FIELD THEORY}%
\author{S.~Mignemi}%1 автор
\affiliation{Dipartimento di Matematica e Informatica, Universit\'a di Cagliari,}%институт
\address{viale Merello 92, 09123 Cagliari, Italy}%адрес
\email{smignemi@unica.it}%e-mail
\affiliation{INFN, Sezione di Cagliari}%
\address{Cittadella Universitaria, Monserrato, Italy}%

%\udk{539} \pacs{71.20.Nr, 72.20.Pa} \razd{\secvii}

\autorcol{S.~Mignemi}

\setcounter{page}{1}%

\begin{abstract}
We review the main features of the relativistic Snyder model and its generalizations.
We discuss the quantum field theory on this background using the standard formalism of
noncommutaive QFT and discuss the possibility of obtaining a finite theory.
\end{abstract}


\maketitle

\section{Introduction}

Since the origin of quantum field theory (QFT) there have been proposal to add a new scale of length to the theory in order to solve the
problems connected to UV divergences.
Later, also attempts to build a theory of quantum gravity have proved the necessity of introducing a length scale, that has been
identified with the Planck length $L_p=\sqrt{\hbar G\over c^3}\sim1.6\cdot 10^{-35}\;{\rm m}$ \cite{Garay}.
A naive application of this idea, like a lattice field theory, would  however break Lorentz invariance.
A way to reconcile discreteness of spacetime with Lorentz invariance was proposed by Snyder \cite{Snyder 1947} a long time ago.
This was the first example of a noncommutative geometry: the length scale should enter the theory through the commutators of
spacetime coordinates.

Noncommutative geometries were however not investigated for a long time, until they revived due to mathematical \cite{Connes} and
physical \cite{ncg} progresses. Their present understanding is based on the formalism of Hopf algebras \cite{Majid}.
In particular, QFT on noncommutative backgrounds has been largely studied \cite{ncg}.
In most cases, a surprising phenomenon, called UV/IR mixing, occurs: the counterterms needed for the UV regularization diverge
for vanishing incoming momenta, inducing an IR divergence.

Noncommutative geometries also admit a sort of dual representation on momentum space in theories of doubly special relativity (DSR)
\cite{Amelino-Camelia}. Here a fundamental mass scale is introduced, that causes the curvature of momentum space \cite{KG},
and the deformation of both the \poi group and the dispersion relations of the particles.
The Snyder model can also be seen as a DSR model, where the \poi invariance and the dispersion relations are undeformed.

As mentioned above, Snyder's idea was almost abandoned with the introduction of renormalization techniques, with the exception of some
Russian authors in the sixties \cite{russi}.
It revived more recently, when noncommutative geometry became an important topic of research. However,
in spite of several attempts using various methods \cite{russi,QFT}, the issue of finiteness of Snyder field theory has not been
established up to now.
Here we review an attempt to investigate this topic using the formalism of noncommutative QFT \cite{You,U2}.

\section{The Snyder model}
The most notable feature of the Snyder model is that, in contrast with most examples of noncommutative geometry,
it preserves the full \poi invariance.
In fact, it is based on the Snyder algebra, a deformation of the Lorentz algebra acting on phase space, generated by positions ${x_\m}$,
momenta ${p_\m}$ and Lorentz generators ${J_\mn}$, that obey the \poi \cor
\be\lb{1}
[J_\mn,J_{\r\s}]=i\big(\y_{\m\r}J_{\n\s}-\y_{\m\s}J_{\n\r}+\y_{\n\r}J_{\m\s}-\y_{\n\s}J_{\m\r}\big),\ee
\be\lb{2}
\quad[p_\m,p_\n]=0,\qquad[J_\mn,p_\l]=i\left(\y_{\m\l}p_\n-\y_{\l\n}p_\m\right),
\ee
together with the standard Lorentz action on position
\be\lb{3}
[J_\mn,x_\l]=i\left(\y_{\m\l}x_\n-\y_{\n\l}x_\m\right),
\ee
and a deformation of the Heisenberg algebra (preserving the Jacobi identities),
\be\lb{heis}
[x_\m,x_\n]=i\b J_\mn,\qquad[x_\m,p_\m]=i(\y_\mn+\b p_\m p_\n),
\ee
where ${\b}$ is a parameter of the order of the square of the Planck length and ${\y_\mn=\diag(-1,1,1,1)}$.
The generators ${J_\mn}$ are realized in the standard way as ${J_\mn=x_\m p_\n-x_\n p_\m}$.

In contrast with most models of noncommutative geometry, the commutators \eqref{heis} are  functions of the phase
space variables: this allows them to be compatible with a linear action of the Lorentz symmetry on phase space.
However, translations act in a nontrivial way on position variables.

It is important to remark that, depending on the sign of the \cc ${\b}$, two rather different models can arise:
\bea\nn
&{\b>0}\qquad\quad&{\rm Snyder\ model}\cr
&{\b<0}\qquad\quad&{\rm anti-Snyder\ model}
\eea
They have very different properties. For example, the Snyder model has a discrete spatial structure and a continuous
time spectrum, while the opposite holds for anti-Snyder.

The subalgebra generated by $J_\mn$ and $x_\m$ is isomorphic to the de Sitter/\ads algebra, and hence
the Snyder/anti-Snyder momentum spaces have the same geometry as de Sitter/\ads spacetime respectively. In fact,
the Snyder momentum space can be represented as a hyperboloid ${\cH}$ of equation
\be\lb{hyper}
\z_A^2=1/\b
\ee
embedded in a 5D space of \coo $\z_A$ with signature $(-,+,+,+,+)$, or equivalently as a coset space $SO(1,4)/SO(1,3)$.

The Snyder \cor are recovered through the choice of isotropic (Beltrami) \coo on ${\cH}$
\be\lb{beltrami}
p_\m=\z_\m/\z_4
\ee
and the identification
\be\lb{identif}
x_\m=M_{\m4},\qquad J_\mn=M_\mn.
\ee
where ${M_{AB}}$ are the Lorentz generators in 5D.
Note that this construction implies that $p^2<1/\b$, and hence the existence of a maximal mass, of the order of the
Planck mass, for elementary particles.
This is a common feature in models with curved momentum space \cite{KG}.

The momentum space of the anti-Snyder model can be represented analogously, as a hyperboloid of equation
\eqref{hyper} with $\b<0$, embedded in a 5D space of \coo {$\z_A$} with signature {$(+,-,-,-,+)$}, or equivalently as
a coset space $SO(2,3)/SO(1,3)$.
Again, anti-Snyder \cor are recovered through the choice of isotropic  \coo \eqref{beltrami}
and the identification \eqref{identif}. An important difference from the previous case is that
the momentum squared is now unbounded. In the following we shall concentrate on Snyder space, but most results hold also for $\b<0$.

\section{Generalizations of the  Snyder model}
The Snyder model can be generalized by choosing different isotropic parametrizations of the momentum space, but maintaining the
identification $x_\m=M_{\m4}$. In this way, eqs.~(\ref{1}-\ref{3}) and the  position \cor still hold, but $[x_\m,p_\n]$  is deformed.
For example, choosing $p_\m=\z_\m$, one obtains \cite{Battisti}
\be
[x_\m,x_\n]=i\b J_\mn,\qquad [x_\m,p_\n]=i\sqrt{1+\b p^2}\,\y_\mn.
\ee
The most general choice that preserves the \poi invariance is \cite{Ivetic}
\be
p_\m=f(\z^2)\z_\m,\qquad x_\m=g(\z^2)M_{\m4}.
\ee

Algebraically, the same models can also be obtained by deforming the Heisenberg algebra as \cite{Battisti,MMMS}
\bea
&&[x_\m,x_\n]=i\b J_\mn\,\q(\b p^2),\cr
&&[x_\m,p_\n]=i\big[\y_\mn\f_1(\b p^2)+\b p_\m p_\n\f_2(\b p^2)\big].
\eea
The function $\f_1$ and $\f_2$ are arbitrary, but the Jacobi identity implies
\be
\q =\f_1\f_2-2(\f_1+\b p^2\f_2)\ {d\f_1\over d(\b p^2)}.
\ee

A different kind of generalization is obtained by choosing a curved spacetime  (de Sitter) background, imposing nontrivial
\cor between the momentum variables,
\be
[p_\m,p_\n]=i\a J_\mn,
\ee
with $\a$ proportional to the cosmological constant. This idea goes back to Yang \cite{Yang}, but was later elaborated
in a more compelling way in  \cite{Kowalski}. We call this generalization Snyder-\des (SdS) model.

The other \cor are unchanged, except that now, by the Jacobi identities,
\bea
[x_\m,p_\n]&=&i\Big(\y_\mn+\a x_\m x_\n+\b p_\m p_\n\nn\\
&&+\sqrt{\a\b}(x_\m p_\n+p_\m x_\n)\Big).
\eea
This model depends on two invariant scales besides the speed of light, that are usually identified with
the Planck mass and the cosmological constant, from which the alternative name name triply special relativity,
proposed in \cite{Kowalski} for this model. It must be noted that, in order to have real structure constants,
both $\a$ and $\b$ must have the same sign.
There are indications that the introduction of $\a$ might be necessary to obtain a
well-behaved low-energy limit of quantum gravity theories \cite{Kowalski}.

An interesting property of the SdS model is its duality for the exchange $\a x\leftrightarrow\b p$ \cite{Guo},
that realizes the Born reciprocity \cite{Born}.
The phase space can be embedded in a 6D space as
${SO(1,5)\over SO(1,3)\times O(2)}$ if $\a,\b>0$, or  as $SO(2,4)\over SO(1,3)\times O(2)$ if  $\a,\b<0$
\cite{Mignemi}.

Alternatively, one can construct the SdS algebra directly from that of Snyder by the nonunitary transformation
\be\lb{sds}
x_\m=\hat x_\m+\l{\b\over\a}\hat p_\m,\qquad p_\m=(1-\l)\hat p_\m-{\a\over\b}\hat x_\m,
\ee
where $\hat x_\m$, $\hat p_\m$ are generators of the Snyder algebra and $\l$ a free parameter \cite{Mignemi}.

\section{Phenomenological applications}
A wide literature considers the phenomenological implications of the nonrelativistic Snyder model,
especially in connection with the generalized uncertainty principle (GUP) \cite{GUP}.
However, here we are interested in the relativistic case, which has obtained much less consideration.
Some consequences are:

- Deformed relativistic uncertainty relations: from the deformed Heisenberg algebra one gets
\be
\D x_\m\D p_\n\ge\ha(\y_\mn+\b\D p_\m\D p_\n).
\ee
The spatial components essentially coincide with those considered in GUP.

- Modification of perihelion shift of planetary orbits \cite{Strajn}: provided that the model is applicable to
macroscopic phenomena, on a \sch backgorund the perhihelion shift gets an additional contribution,
$\d\theta=\d\theta_{rel}\left(1+{5\over3}\b m^2\right)$,
where $m$ is the mass of the planet. This correction clearly breaks the equivalence principle at Planck scales.

- DSR-like effects  \cite{Rosati}: no effects of time delay of cosmological photons occur, contrary to other
models derived from noncommutative geometry \cite{Mercati},
but some higher-order effects are still present.

\section{Hopf algebras}
In the study of noncommutative models an important tool is given by the Hopf algebra formalism \cite{Majid}, especially
in relation with QFT.

Since in noncommutative geometry spacetime \coo are noncommuting operators, the composition of two plane waves
$e^{ip\cdot x}$ and  $e^{iq\cdot x}$
gives rise to nontrivial addition rules for the momenta, denoted by $p\oplus q$, that are described by the coproduct structure
of a Hopf algebra, $\D(p,q)$. The addition law is in general noncommutative.

Moreover, the opposite of the momentum is determined by the antipode of the Hopf algebra,
$S(p)$, such that $p\oplus S(p)=S(p)\oplus p=0$.

The algebra associated to the Snyder model can be calculated (classically) using the geometric representation
of the momentum space as a coset space mentioned above and calculating the action of the group multiplication on it
\cite{Girelli}.

Alternatively, one can use the algebraic formalism of realizations \cite{Battisti}:
a realization of the noncommutative coordinates $x_\m$ is defined in terms of coordinates
$\x_\m$, $p_\m$   that satisfy  canonical \cor
\be
[\x_\m,\x_\n]=[p_\m,p_\n]=0,\qquad[\x_\m,p_\n]=\y_\mn,
\ee
by assigning a function $x_\m(\x_\m,p_\m)$ that satisfies the Snyder commutation relations.

The $x_\m$ and $p_\m$ are now interpreted as operators acting on function of $x_\m$, as
$$\x_\m\rhd f(\x)=\x_\m f(\x),\qquad p_\m\rhd f(\x)=-i\de f(\x)/\de\x_\m.$$
In particular, it is easy to show that the most general realization of the Snyder model is given by \cite{MMMS}
\be
x_\m=\x_\m +\b\,\x\bdot p\,p_\m+\b \c(\b p^2)\,p_\m,
\ee
where the function $\c(\b p^2)$ is arbitrary and does not contribute to the commutators, but takes into account ambiguities
arising from operator ordering of $\x_\mu$ and $p_\mu$.

In general, it can be shown that for any noncommutative model, \cite{Svrtan}
\be
e^{ik\cdot x}\rhd e^{iq\cdot\x}=e^{i\cP(k,q)\cdot\x+i\cQ(k,q)},
\ee
where the functions $\cP_\m$ and $\cQ$  can be deduced from the realization.
Moreover,
\be
e^{ik\cdot x}\rhd1=e^{i\cK(k)\cdot \x+i\cJ(k)},
\ee
with ${\cK_\m(k)\id\cP_\m(k,0)}$ and ${\cJ(k)\id\cQ(k,0)}$.
The generalized addition of momenta is then given by
\be
k_\m\oplus q_\m=\cD_\m(k,q),
\ee
with $\cD_\m(k,q)=\cP_\m(\cK^\mo(k),q)$, and the coproduct is simply
\be
\D p_\m=\cD_\m(p\otimes1,1\otimes p).
\ee
Note that ${\cD_\m}$ is independent of ${\c}$.
Moreover, the antipode  $S(p_\m)$,  is  $-p_\m$ for all (generalized) Snyder models.

A fundamental property of the Snyder addition law is that it is  nonassociative. Hence the algebra is
noncoassociative, so strictly not a Hopf algebra.

For the calculations, it is also useful to define a star product, that  gives a representation of the product of functions of the noncommutative
coordinates $x$ in terms of a deformation of the product of functions of the commuting coordinates $\x$.
In particular, from the previous results one can calculate the star product of two plane waves:
\be
e^{ik\cdot\x}\star e^{iq\cdot\x} = e^{i\cD(k,q)\cdot\x+i\cG(k,q)},
\ee
where
\be
\cG(k,q)=\cQ(\cK^\mo(k),q)-\cQ(\cK^\mo(k),0).
\ee

We consider now a Hermitean realization of the Snyder \cor
\bea\lb{herm}
x^\m&=&\x^\m+{\b\over2}\left(\x\bdot p\, p^\m+p^\m p\bdot\x\right)\cr&=&\x^\m+\b\,\x\bdot p\,p^\m-{5i\over2}\b\, p^\m.
\eea
The request of Hermiticity will be important for the field theory.
We get
\bea
&\cD_\m(k,q)=&{1\over1-\b k\bdot q}\Bigg[\left(1+{\b\,k\bdot q\over1+\rad}\right)k_\m\cr
&&\qquad\qquad\qquad+\rad\, q_\m\Bigg],\cr
&\cG(k,q)=&{5i\over2}\ln\left[1-\b\,k\bdot q\right].
\eea
and hence
\be\lb{starp}
e^{ik\cdot\x}\star e^{iq\cdot\x}={e^{i\cD(k,q)\cdot\x}\over(1-\b\,k\bdot q)^{5/2}}.
\ee

\section{QFT in Snyder space}

Let us consider a QFT for a scalar field $\f$ on a Snyder space.
Usually, field theories in noncommutative spaces are constructed by continuing to Euclidean signature
and writing the action in terms of the star product \cite{ncg}.

In fact,
the action functional for a free massive real scalar field $\phi(x)$ can be defined through the star product as
\cite{MMMS}
\be
S_{\rm free}[\phi]=\ha\int d^4\x\,(\de_\m\phi\star\de^\m\phi+m^2\phi\star\phi).
\ee
The star product of two real scalar fields $\phi(\x)$ and $\q(\x)$ can be computed
by expanding them in Fourier series,
\be
\phi(\x)=\int d^4k\,\tilde\phi(k)e^{ik\cdot\x}.
\ee
Then, using \eqref{starp},
\bea
&&\int d^4\x\ \q(\x)\star\phi(\x)=\int d^4\x\int d^4k\,d^4q\ \tilde\q(k)\,\tilde\phi(q)\ \cr
&&\times\,e^{ik\cdot\x}\star e^{iq\cdot\x}=\int d^4k\,d^4q\ \tilde\q(k)\,\tilde\phi(q)\ {\d^{(4)}\big(\cD(k,q)\big)\over(1-\b\,k\bdot q)^{5/2}}.\cr&&\eea
But
\bea
\d^{(4)}\big(\cD(k,q)\big)&=&{\d^{(4)}(q+k)\over\left|\det\left({\de\cD_\m(k,q)\over\de q_\n}\right)\right|_{q=-k}}\cr
&=&(1+\b k^2)^{5/2}\d^{(4)}(q+k).
\eea
The two $(1+\b k^2)^{5/2}$ factors cancel and then \cite{MMMS},
\be
\int d^4\x\ \q(\x)\star\phi(\x)=\int d^4\x\ \q(\x)\,\phi(\x).
\ee
This is called cyclicity property, and occurs also in other noncommutative models; it follows that the free theory is identical to the
commutative one,
\be
S_{\rm free}[\phi]=\ha\int d^4\x\left(\de_\m\phi\,\de_\m\phi+m^2\phi^2\right).
\ee
The propagator is therefore the standard one
\be
G(k)={1\over k^2+m^2}.
\ee
Notice that the cyclicity property is a consequence of our choice of a Hermitian representation for the operator $x$, and can be related
to the choice of the correct measure in the curved momentum space.

The interacting theory is much more difficult to investigate.
Several problems arise:

- The addition law of momenta is noncommutative and nonassociative, therefore
one must define some ordering for the lines entering a vertex and then take an average.

- The conservation law of momentum is deformed at vertices, so loop effects may lead to nonconservation of momentum in a propagator.

For example, let us consider the simplest case, a $\phi^4$ theory \cite{You}
\be\lb{inter}
S_{\rm int}=\l\int d^4\x\ \phi\star(\phi\star(\phi\star\phi)).
\ee
The parentheses are necessary because the star product is  nonassociative. Our definition fixes this ambiguity, but other choices are possible.

With this choice, the 4-point vertex function turns out to be
\be
\begin{split}
G^{(0)}(p_1,p_2,p_3,p_4)=\sum\limits_{\sigma\in S_4}&\delta\Big(\cD_4\big(\sigma(p_1,p_2,p_3,p_4)\big)\Big)\\
&\times\, g_3\big(\sigma(p_1,p_2,p_3,p_4)\big),\nn
\end{split}
\ee
where
\be
\cD_4(q_1,q_2,q_3,q_4)=q_1+\cD(q_2,\cD(q_3,q_4)),
\ee
\be
g_3(q_1,q_2,q_3,q_4)=e^{i\cG(q_2,\cD(q_3,q_4))}e^{i\cG(q_3,q_4)},
\ee
and $\s$ denotes all possible permutations of the momenta entering the vertex.

With the expressions of the propagator and the vertex one can compute Feynman diagrams.
For example, the one-loop two-point function depicted in fig.~1 in position space is given by
\begin{equation}
\begin{split}
G^{(1)}&(\x_1,\x_2)=\cr&-\frac{1}{2}\frac{\lambda}{4!}\int d^4 p_1d^4 p_2d^4 \ell
\frac{e^{ip_1\x_1}}{p_1^2+m^2}\frac{e^{ip_2\x_2}}{p_2^2+m^2}\frac{1}{\ell^2+m^2}\\
&\sum\limits_{\sigma}\delta\Big(D_4\big(\sigma\big(p_1,p_2,\ell,-\ell\big)\big)\Big)\
g_3\Big(\sigma\big(p_1,p_2,\ell,-\ell\big)\Big).
\end{split}
\end{equation}
\begin{figure}% figure* for wide figure, [h] [!] to change the placement
\vskip1mm
\includegraphics[width=\column]{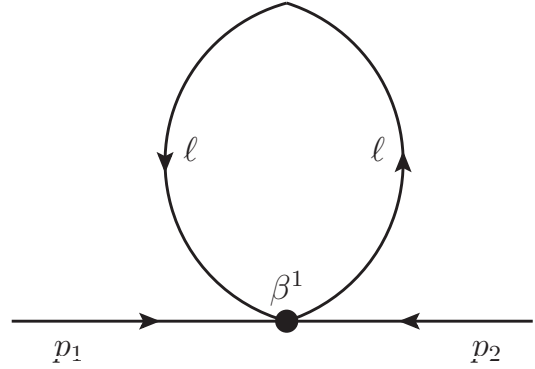}
\vskip-3mm\caption{One-loop two-point function.}
\end{figure}
To evaluate the diagram, one must consider the 24 permutations of the momenta entering the vertex.
Among  these, only 8 conserve the momentum (i.e. $p_1=-p_2$), while the remaining 16 do not.

At the linear level in $\b$ the calculation can be done explicitly, showing stronger divergences than in the commutative theory \cite{You}.
However, the effects of momentum nonconservation cancel out.

Attempting instead a calculation at all orders in $\b$, not all diagrams can be explicitly computed \cite{U2}.
It can be shown, however, that the divergences are suppressed with respect to the
noncommutative theory and there are even indications that the integrals might be finite, at least for the interaction \eqref{inter}.

If instead, renormalization is necessary, the phenomenon of UV/IR mixing could still be present, as in other noncommutative models
\cite{ncg}.
We recall however that a model that avoids this problem in Moyal theory was proposed by Grosse and Wulkenhaar \cite{Grosse} (GW model).
Its main characteristic is that, besides the kinetic and interaction terms, its action also contains a term proportional to $\f\,x^2\f$.
A similar mechanism can be recovered in Snyder theory by considering a curved background (SdS model) \cite{Franchino}.

In fact, using the relation \eqref{sds} between the SdS and Snyder algebra with $\l=0$,
and the realization \eqref{herm} of the Snyder algebra, the action can be reduced, at zeroth order in $\a$, $\b$, to
\be\lb{gw}
S_{\rm free}=\int d\x^4\f\left[p^2+\frac{\alpha}{\beta}\,\x^2+m^2+O(\a,\b)\right]\f,
\ee
The action \eqref{gw}  is identical to that of the free GW model.
One may therefore hope that also in this case the IR divergences are suppressed  and one can obtain a renormalizable theory.

\section{Conclusions}
We have reviewed the present status of research on the Snyder model, the earliest example of noncommutative geometry,
and the only one that preserves Lorentz invariance.
In particular, we concentrated on the definition of a quantum field theory defined in accord with the standard noncommutative
formalism, and the issue of renormalizability. It turns out that, although an exact calculation has not been performed in full,
there is good evidence of renormalizability and absence of UV/IR mixing, at least in the SdS model.

\end{document}